# A Novel Approach for Handling Misbehaving Nodes in Behavior-Aware Mobile Networking

Kanad Basu Subrata Mitra Srishti Mukherjee Weixun Wang
Department of Computer and Information Science and Engineering
University of Florida, Gainesville, FL

## I. Motivation

Profile-cast is a service paradigm within the communication framework of delay tolerant networks (DTN)[1] Instead of using destination addresses to determine the final destination, the routing protocol uses a profile-based forwarding approach. Thus, profile-cast can be considered a special case of behavioral networks. With the rise in popularity of various wireless networks, the need to make wireless technologies robust, resilient to attacks and failure becomes mandatory. One issue that remains to be addressed in behavioral networks is node co-operation in forwarding packets. Nodes in a behavioral network might behave selfishly (due to bandwidth preservation, energy /power constraints) or maliciously by dropping packets or not forwarding them to other nodes based on profile similarity. These detrimental attitudes of nodes are more prominent when each node is autonomous. In both cases the net result is degradation in the performance of the network.

It is our goal to investigate whether and how the performance of the behavioral network can be improved by employing self-policing scheme that would detect node misbehavior and then decide how to tackle them in order to ensure node-co-operation.

This will try to ensure that the overall performance does not fall below a certain limit. We need to first determine the effect selfish or malicious nodes have on the overall network. Various existing self-policing techniques [2], [3], [4], which are in use in mobile adhoc networks will be tried on the behavioral network, using some form of simulation. Those results will show whether these schemes are applicable in behaviriol networks as well. If none of these approaches work adequately, a novel scheme will be proposed to solve the problem. We now enumerate the various challenges in applications of self-policing techniques to behavioral networks.

1) The main problem associated with this area is the nature of the forwarding protocol. Absolutely no information is stored about which path the data packet took to reach the destination. Hence it is hard to detect which node dropped the packet.
2) Secondly The delay tolerant nature of the network, makes it difficult to have a distributed reputation based system [2], where each node can have tables containing the behavior ratings of each other node.
3) As with ad-hoc networks, the nodes have bandwidth and energy constraints.
4) Nodes might misbehave individually as well as in a group for a collusion, thereby complicating misbehavior detection.

## II. Related Work

Profile-Cast, a novel adhoc networking concept introduced in [1], leverages the behavioral patterns of mobile network users for delivering messages to a sub-group of users as defined by their profiles (e.g., interest, social affiliation, etc.). A total cooperation is needed from all the nodes in order to maintain the overall performance of the network. If the nodes are autonomous, they really have no incentive to cooperate, that is, forward messages destined to other nodes, since forwarding utilizes some physical resources (like battery power, bandwidth, etc.) and these are generally scarce. To ensure node co-operation it is necessary to propose some kind of incentive scheme. As described by [5], an "Analysis" phase is needed to design an incentive based protocol.

- Analysis: The engineer analyzes and adjusts the cooperation protocol that requires an incentive scheme. The analysis phase comprises of checking what type of cooperation exists between the nodes in the network. Further, the cost of such cooperation for each entity is chosen and adjustments are made to render the behavior more perceptible. This phase is allowed to obtain a theoretical knowledge of what needs to be used to enhance cooperation incentives among the nodes.

In the analysis phase we first should propose a trust model so that we have a reference or starting point. Zhaoyu Liu et al. [6] introduces a trust model for mobile ad hoc networks. Initially each node is assigned a trust level. The nodes neighboring to a node exhibiting suspicious behavior initiate trust reports. These trust reports are propagated through the network using one of our proposed methods. A source node can use the trust levels it establishes for other nodes to evaluate the security of routes to destination nodes. Using these trust levels as a guide, the source node can then select a route that meets the security requirements of the message to be transmitted. This paper demonstrates important concepts for establishing a collaborative, dynamic trust model and for using this model as an example to enhance the security of message routing in mobile ad hoc networks. Approaches to incentive scheme include:-

1) Reputation based scheme
2) Credit Based scheme
3) Game-theoretic approach

Buchegger et al. [2] proposed a fully distributed reputation system which make advantage of both first-hand and second-hand reputation information. It can effectively cope with false shared observations. In their later work, Buchegger et al. [7] described general features and design considerations for reputation-based self-policing techniques in mobile ad-hoc networks. SPRITE, a simple cheat proof credit based system was proposed by [3]. They assumed the presence of a centralized Credit Clearance System(CCS), which denotes credits to individual nodes in the network. The basic philosophy remains that whenever a node is forwarding someone else's packet, it is gaining some credit. On the other hand, a node, while trying to send it's own packet looses credit. Thus, a credit balance is maintained so that a node is forced to forward other's packets in order to send it's own. Srinivasan et al. [4] first used a game theory based approach to ensure node co-operation in adhoc networks by trying to optimize power of a node to its throughput. The work is basically aimed to address the issue of selfish nodes. Profile-Cast falls in the paradigm of Delay Tolerant Networks (DTN). Zhu et al [8] have proposed a credit-based incentive scheme to foster co-operation among nodes in a DTN. An incentive aware routing was proposed for DTN networks by Upendra Shivade et al. [9], on the lines similar to Srinivasan et al. [4] which uses pair-wise generous tit-for-tat. Another approach based on game-theory has been proposed by L Buttyan et al. [10] for delay tolerant networks. Our work is different from the existing ones in the inherent nature of the Profile Cast network, and the nature of the trust model which we are going to build. Section III will describe the main features of the project.

## III. Proposed Research

### A. Problem Statement

Profile-cast has two main communication modes using behavioral profiles. One is CSI: Target mode (CSI:T) and another is CSI: Dissemination mode (CSI:D)[1]. CSI:T can be used when the target profile is in the same context as the behavioral profile and CSI:D provides a more generic option when the target profile is irrelevant to the context of the behavioral profile. In other words, for example, if we use mobility as the profile, nodes with similar profile tend to be physically close to each other. However, in CSI:D, they could be scattered in the behavioral space.

CSI:T has two main phases: a) gradient ascend phase and b) group spread phase. In the gradient ascend phase, it sends only one copy of the message (i.e. only one thread) which is delivered to another node which has better profile match with the target profile[1]. After that, the responsibility of further delivering the message is given to that node. This mechanism achieves less than 84% overhead of the delay-optimal strategy (which generates as many copies as needed to achieve the lowest delay for each node) without much degradation in the delivery ratio. CSI:D attempts to send the message to the intended receivers utilizing the small world character of the encounter graph [1]. The sender starts as the first message holder and then adds additional message holders who are very dissimilar in their behavioral profiles, to achieve low coverage overlaps. Each message holder keeps a list of behavioral profiles of its known other message holders.

[1]The misbehaving node might be a concern here, since this *one* packet might be dropped. A better strategy to make it more robust is to forward some more packets to a group of nodes with better profile matches

We can see both of these mechanisms depend on certain kind of trusts among all the nodes. A trust model for profile-cast needs to be defined. Both CSI:T and CSI:D assume that nodes carrying the message to the intended recipients are trustworthy and thus will not misbehave. But in reality some nodes may refuse to forward the packet (may be due to physical limitations like power, bandwidth etc.). Malicious nodes silently dropping the packet and lying about it because they are selfish or they want to disrupt the message forwarding mechanism. There can be individual nodes that are misbehaving or can be a group of colluding nodes in-order to disrupt the system.

In CSI:T, during the initial phase of gradient ascend, there will be only a few nodes with low profile matching with the target. Therefore, the sending node will have very limited choices to select the next sender. Since the matching with the target profile is very low, it is most likely that those nodes will not be very trustworthy as they may not be interested/acquainted about the target profile. It is possible when nodes with similar profile are under the same authority/interest party. Therefore, they may not be willing to use their resource to forward the messages from other groups. Since there are only a small number of packets at this stage, the impact on the delivery ratio will be significant. Similarly, in CSI:D, new message holders are chosen from those who have very dissimilar profiles. It may lead to selection of misbehaving nodes as the next message holder. Note that in CSI:D there will be multiple message holders so the impact of misbehavior may not be as severe as in CSI:T.

We plan to address the following problems:

1) We first need to analysis how much impact will be made on the message delivery mechanism (both for CSI:T and CSI:D) if we introduce malicious nodes who may misbehave either as individuals or a group. Observations need to be made on all the evaluation metrics before/after injecting misbehaving nodes.
2) Profile-cast when used in a Delay Tolerant Network renders early detection of misbehavior or malicious activities difficult because, in most of the cases, the sender has no way to get an acknowledgement from the receiver. Moreover, it is possible that the acknowledgement itself is false because the receiving node is malicious. Therefore, one of the main problems is how to detect node's misbehavior in a behavior-aware mobile network.
3) There are a number of existing self-policing mechanisms for both ad-hoc networks as well as DTN. Specifically, they can fall into the following classifications: incentive/credit-based system, reputation systems and game-theory based approaches. The problem remains to be answered is that whether we can apply these techniques directly or a hybrid approach in our scenario (e.g. profile-cast) to achieve satisfactory results. Particularly, if a reputation-based approach is applied, we also need to build an efficient trust model.
4) Trustworthiness of a misbehaving node may vary de-

pending on the target profile. During gradient ascend phase in CSI:T, a node may fall on the gradient of multiple targeted profiles. Since we have very few choices during this initial stage, if we try to avoid (or blacklist) some nodes after knowing that they dropped packets (or showed other misbehavior), the delay may increase significantly. In the worst case, we may end up with transmission failure due to incapability in finding matching nodes. Moreover, misbehaving to one certain kind of profiles may not mean that it will also misbehave for other profiles (for example, its own matching profiles). Therefore, we need to make a trade-off here between reducing the overhead and maintaining the delivery ratio/delay bounds. It will be a major challenge for us to incorporate all these scenarios and come up with a novel self-policing approach for behavior-aware networks.

### B. *Evaluation Metrics*

The evaluation metrics we would consider including:

1) Message delivery ratio (percentage of the messages which are received in the destination node);
2) Message transmission delay;
3) Transmission overhead (total number of messages transmitted in the delivery process);
4) Storage and Computation overhead (how much intermediate information we need to store);
5) Detection time of misbehaving nodes.

### C. *Investigated Parameter Space*

We will investigate parameters which can affect any of the above mentioned evaluation metrics. In our experiments, we plan to vary those parameters to see how the corresponding metrics change. Specifically, the parameters and their value spaces include (but are not limited to):

1) Percentage of misbehaving nodes in the network which varies from 10% to 100% in one step of 10%; the distribution can be made random or following some statistical patterns;
2) Packet dropping probability of misbehaving nodes which varies from 10% to 100% in one step of 10%;
3) Percentage of untrustworthy nodes in the network, if reputation-based approach is used, which varies from 10% to 100% in one step of 10%;
4) Data-rate of each node (packets sent per second) which varies from 1 to 10 in one step of 1;

### D. *Methodology*

We plan to explore the impact of misbehaving nodes on behavior-aware mobile networks, analyze the applicability of existing techniques and evaluate the effectiveness of our approach using simulation. Possible candidates include NS2 and GloMoSim. We may also be ready to implement our own system prototype as well as the simulator.

### E. *Scenarios*

Since we have real traces collected from University campuses, we would focus on campus like scenarios. We can use mobility behavior as each node's profile. When some nodes misbehave, they may be only interested in forwarding packets to those nodes which have similar mobility behaviors (e.g. student have similar course schedule) since they may have common interests (e.g. homeworks etc.). We believe it is a reasonable and practical scenario.

## IV. Malicious Nodes in Profile-Cast

In this section, we investigate the impact on various evaluation metrics from malicious nodes. First, we describe the overflow of our work. Second, we introduce various parameters that we use to model different scenarios with malicious nodes. Finally, we show the experiment results on how malicious nodes will affect the performance of profile-cast, which motivates our approach for handling misbehaving nodes.

### A. *Overview*

Figure 1 shows the overflow of simulating profile-cast with malicious nodes. The mobility trace files are parsed to generate two kind of files for each node: association matrix and encounter traces. The former one contains the information of each node's association with each location during each time unit (e.g. 1 day in our case). An entry would be, for example, one node spends 30% of the time in the gym on November 2nd, 2007. In other words, it describes each node's mobility preferences which is used as the behavioral profile in our case. The later one stores all the encounters with other nodes for each node. It is assumed that packet transmission between nodes will only happen when they encounter at the same location. In this case, it means they are associated with the same AP. The association matrices will then go through a singular value decomposition (SVD) process which is done by MATLAB. The output, S and V matrices, will be used in calculating the similarity between any pair of nodes in terms of their behavioral profiles. The similarity information will be used during profile-cast.

We randomly generated the target profiles of packets as well as their sender nodes and time. The malicious node information gives values of those parameters which will be discussed in Section IV-B. The target profiles, malicious node information, encounter traces along with S/V matrices are feed into the profile-based simulator. Note that since we assume misbehaviors will only happen during the gradient ascend phase of profile-cast, the simulator will simulate the gradient ascend phase and regards the scenario in which the current message holder's similarity to the target profile reaches the flooding threshold (denoted as $\delta$) as a successful delivery. After the threshold is reached, the gradient ascend phase will stop and it will start flooding the packet. The details about profile-cast can be found in [1].

We will first observe how the performance, in terms of delivery ratio and average transmission delay, will be affected by introducing malicious nodes in this section. Then, our credit-based scheme and reputation-based scheme will be incorporated with the simulator to handle those misbehaving nodes in profile-cast.

### B. *Misbehaving Node Parameters*

We use three parameters to describe the malicious node behaviors in profile-cast:

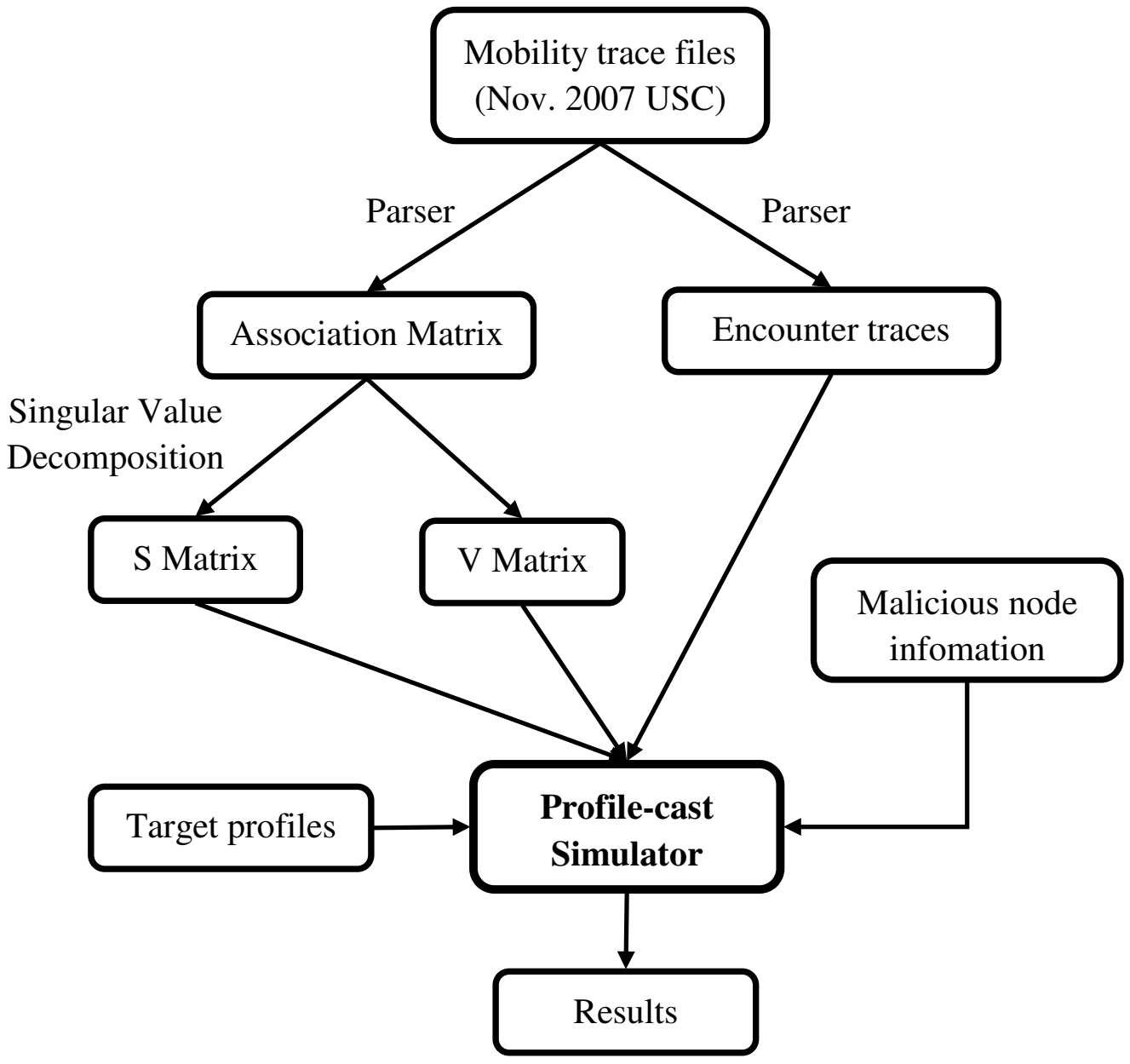

Fig. 1. Overview of profile-cast simulation.

1) Percentage of malicious nodes in all the nodes ($P_1$): We vary the number of malicious nodes in the network from 0% to 100% in one step of 10%. With 0%, it means there is no malicious node while 100% means all the nodes will misbehave with a certain possibility during gradient ascend phase of profile-cast.
2) Lower bound of the randomly generated possibility that one malicious node will actually misbehave during a packet transmission ($P_2$): For each malicious node, it does not always misbehave when a packet is sent to it. In order to model the realistic scenario where malicious node will only misbehave under certain circumstances (e.g. low battery), we use a randomly generated possibility to measure the chance by which the malicious nodes will actually misbehave. $P_2$ acts as the lower bound (the upper bound is 100%) of this randomly generated possibility and varies from 0% to 100% in one step of 10%.
3) Threshold of the similarity to target profile lower than which the malicious node will misbehave ($P_3$): We assume that malicious nodes tend to drop the packet with a target profile dissimilar to themselves. In other words, if the target profile is similar enough to itself, the misbehaving node will forward the packet instead of dropping it. $P_3$ acts as this similarity threshold for making the forwarding decisions and ranges from 10% to $\delta$ in one step of 10%. Note that $\delta$ is the flooding threshold in profile-cast.

Clearly, these three parameters will affect the performance of profile-cast, as shown by our experiment results in next section.

### C. Results

We have used USC November 2007 traces from access points. Which had originally 135 access points and more than 32000 unique MAC ids. At first we filtered some MAC ids based on regularity criteria, i.e. we chose only those MAC ids who logged in more than 40 times in a month or there cumulated log in duration was above a threshold. In this way we identified 3426 unique nodes. For ease of illustration purpose, we randomly generated 1000 packets which are all able to be transmitted to the destination in the original profile-cast. In other words, if there is no malicious node, the delivery ratio is 100%. The flooding threshold $\delta$ is set to 80%.

We investigate the impact on delivery ratio from varying the three parameters described in Section IV-B. Since no misbehavior handling scheme is employed, which means the packet transmission terminates whenever the packet gets dropped, average delay or number of hops is not relevant thus not shows here[2].

*1) Varying Number of Malicious Nodes:* We set the misbehaving possibility lower bound ($P_2$) and the similarity threshold of misbehaving ($P_3$) to 50% and 20%, respectively. Figure 2 shows the result. I can be observed that, as expect, misbehaving nodes will have severe impact on profile-cast's performance. 10% of malicious nodes will lead to 17% packet transmission failures (delivery ratio of 83%). If all the nodes are malicious nodes (with misbehaving possibilities randomly generated within [50%,100%]), the delivery ratio drops below 30%.

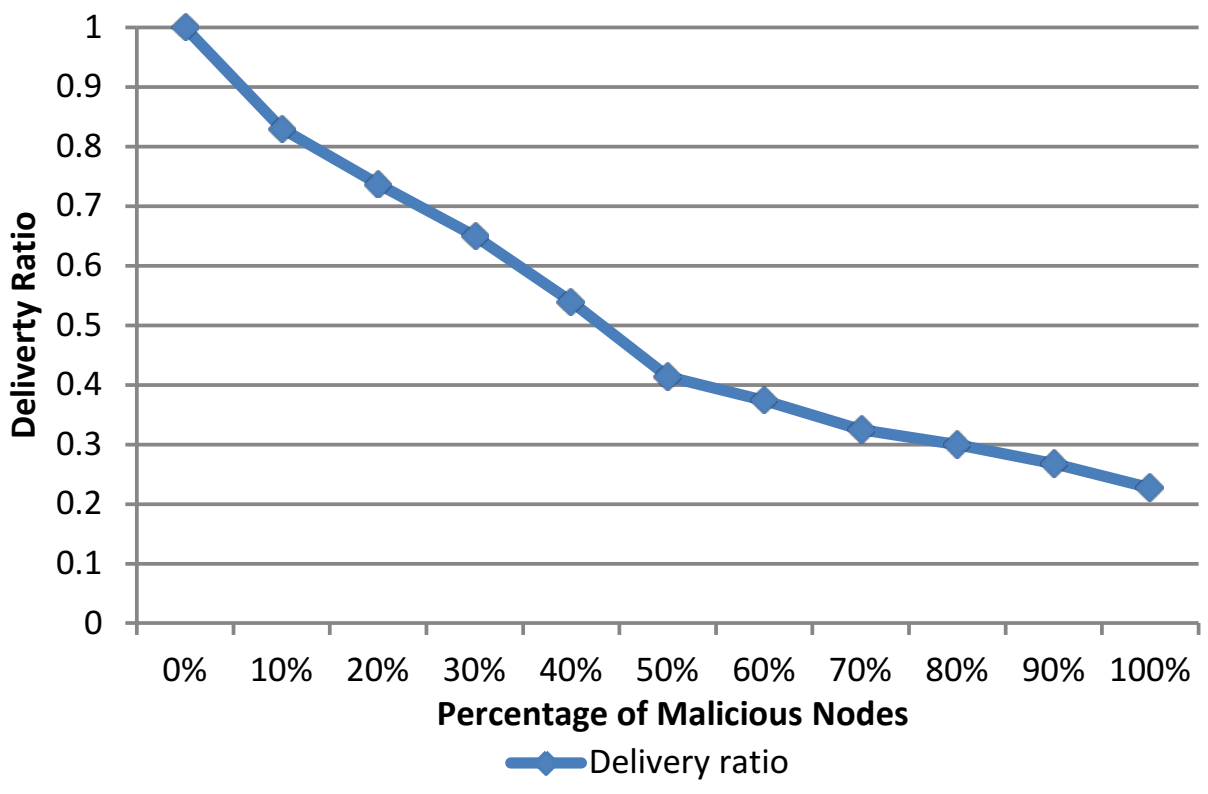

Fig. 2. Impact on delivery ratio from malicious nodes.

*2) Varying Misbehaving Possibility Lower Bound:* In the next set of experiments, we vary $P_2$ as described above. $P_1$ and $P_3$ are set to 30% and 40%, respectively. As shown in Figure 3, when each malicious node has higher chance to actually misbehave at runtime, the delivery ratio will decrease. Note that, however, the degradation is not as drastic as from the number of malicious nodes since $P_2$ acts as the lower bound of the misbehaving possibility.

*3) Varying Similarity Threshold of Misbehaving:* We also investigate the impact on delivery ratio from various similarity threshold to the target profile lower than which the malicious node will misbehave ($P_3$). With $P_1$ and $P_2$ having a value of 40% and 50%, Figure 4 demonstrates this affect. We observe that higher misbehaving threshold will lead to lower delivery ratio since malicious node will drop the packet more likely

[2] In fact, they are actually decreasing with increasing number of malicious nodes due to early-terminated packet transmissions.

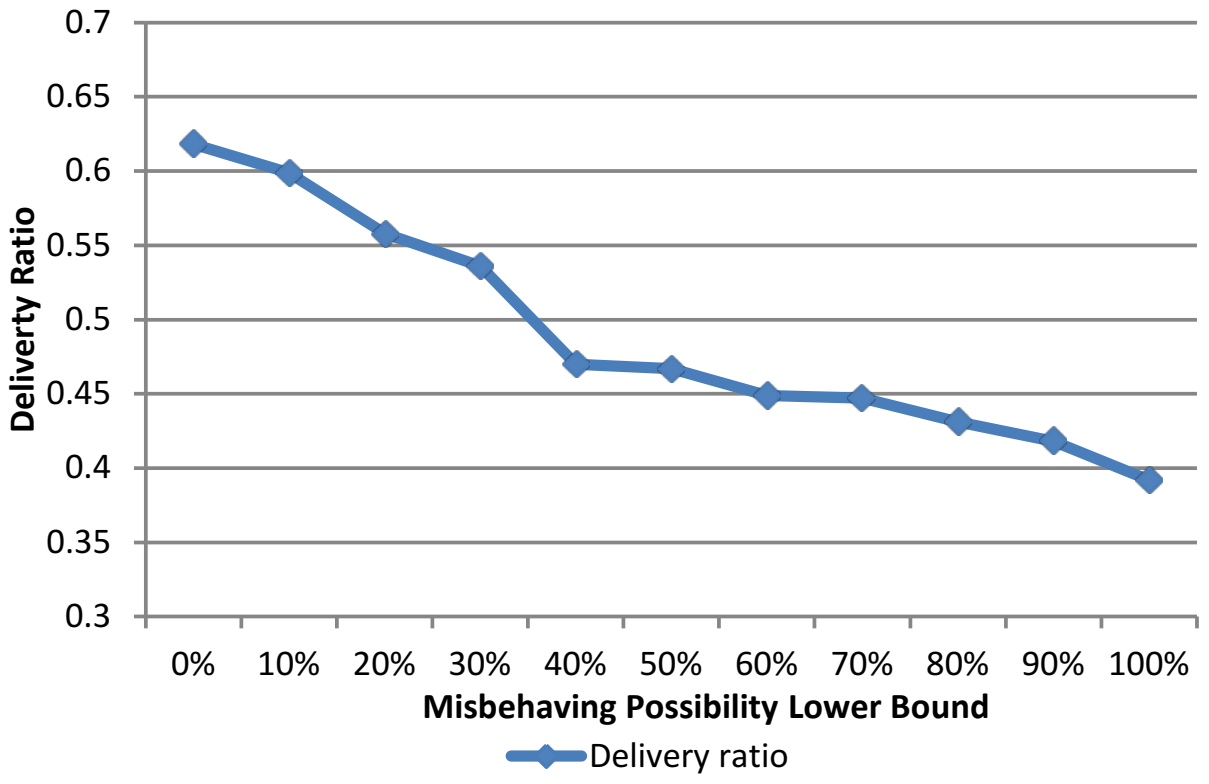

Fig. 3. Impact on delivery ratio from varying misbehaving possibility lower bound.

(i.e. its similarity to the target is lower than the threshold). The degradation in delivery ratio is especially significant when $P_3$ drops from 10% to 30% since it causes most of the packets being dropped during the every early stage of gradient ascend when the packet holder's similarity is low.

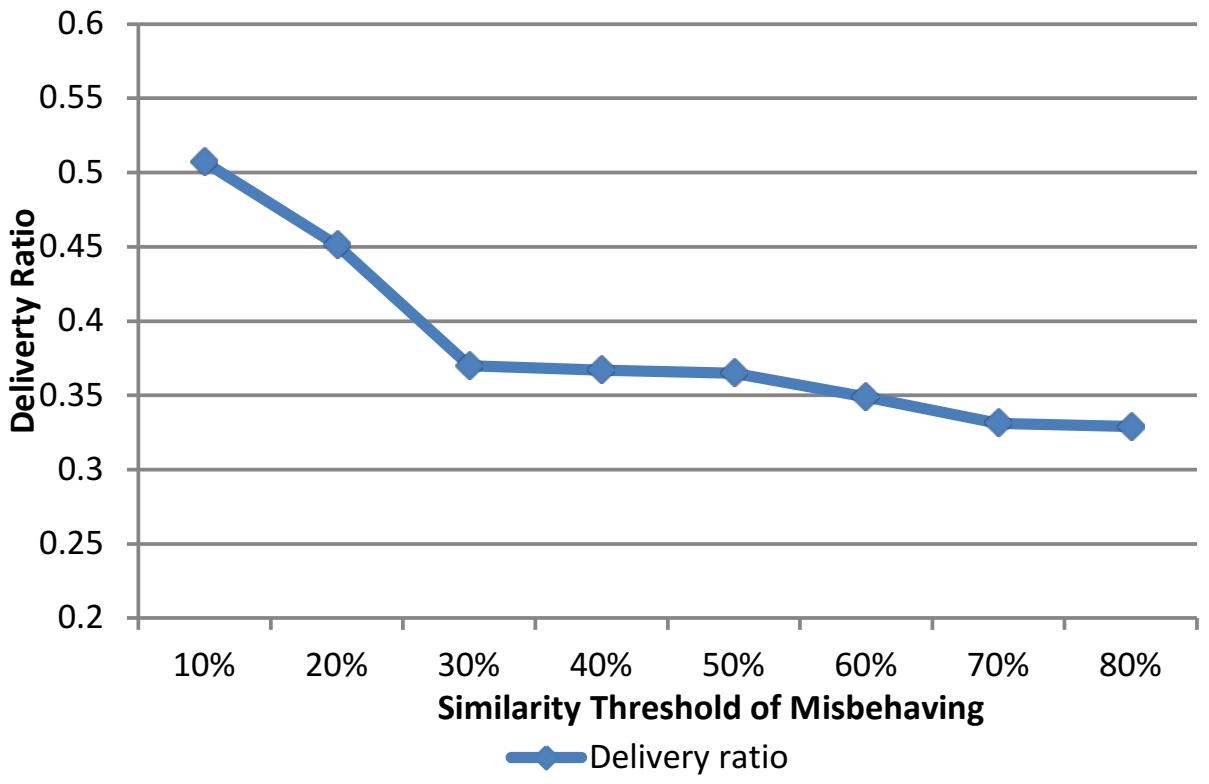

Fig. 4. Impact on delivery ratio from varying similarity threshold of misbehaving.

## V. A Simple Approach – Retransmission

The most intuitive approach to handle malicious nodes in profile-cast is to employ a retransmission scheme. In other words, the current packet holder retransmit the packet to the next encountered node whenever it detects that the last node it sent to misbehaves and drops the packet. Retransmission, clearly, requires certain assumption that each node is able to detect the misbehaviors on its next hop. We feel that it is reasonable in profile-cast since the packet transmission only happens when the two nodes are in the same location (AP). Therefore, we could let the node which receives the packet send an acknowledgement to the sender when it decides to forward the packet. We further assume that misbehaving nodes will not send any acknowledgement. This makes it possible for the sender to setup a timer and reselect the next-hop node again whenever there is a timeout. The process continues until the current packet holder receives the acknowledgement indicating that the packet will get transmitted.

Since malicious nodes do not misbehave all the time, it is possible that it will forward the packet normally in the next encounter with the sender. Therefore, initially, we do not block the malicious node which drops the packet. In other words, if the next encountered node is the same malicious node, the sender will still try to forward the packet to it. We first look at the effect of using retransmission without isolating malicious nodes, after which we will examine whether it is beneficial to block malicious nodes.

Figure 5 illustrates that retransmission is able to improve the delivery ratio significantly. We vary the number of malicious nodes ($P_1$) and set $P_2$ and $P_3$ to 80% and 60%, respectively. The timer's length is set to 1000 seconds. It can be observed that the delivery ratio is increased absolutely by 45%. However, on the other hand, as shown in Figure 6, the average delay due to retransmission is also increased significantly by 1.5 times on average. Clearly, it is a tradeoff which has to be made by the designer between delivery ratio and average delay.

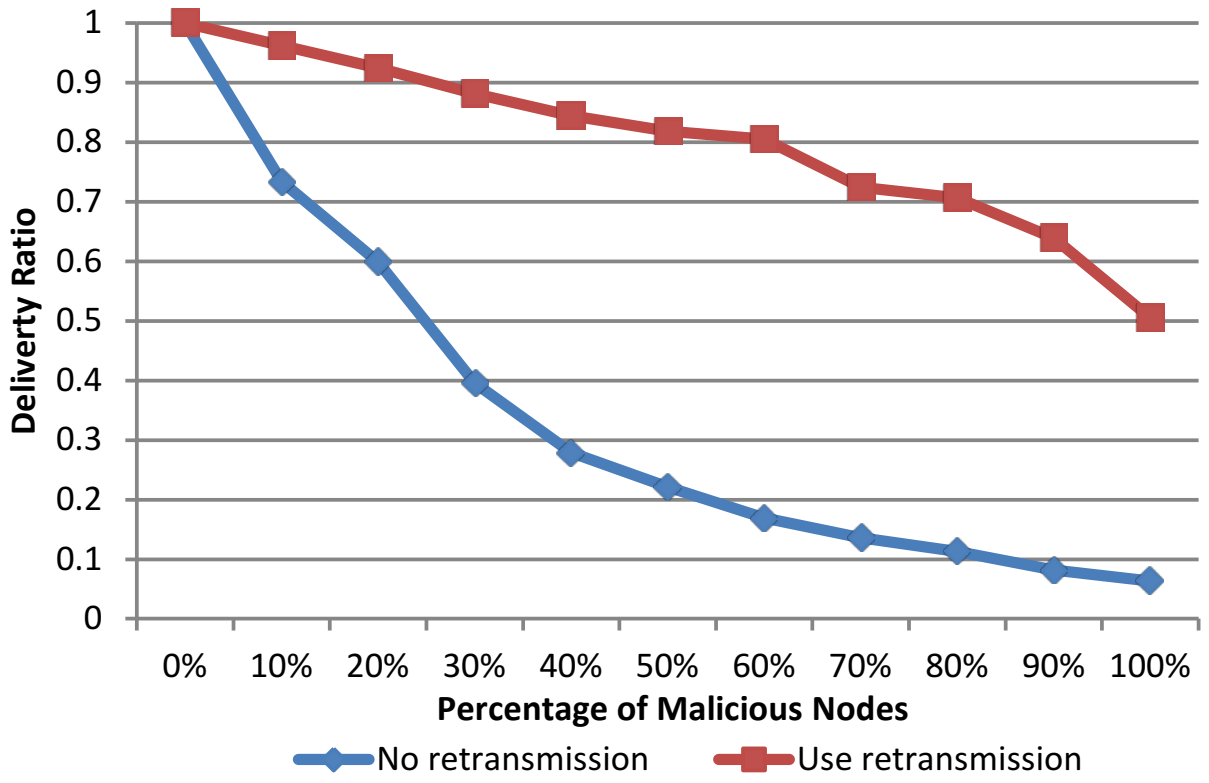

Fig. 5. Delivery ratio improvement using retransmission.

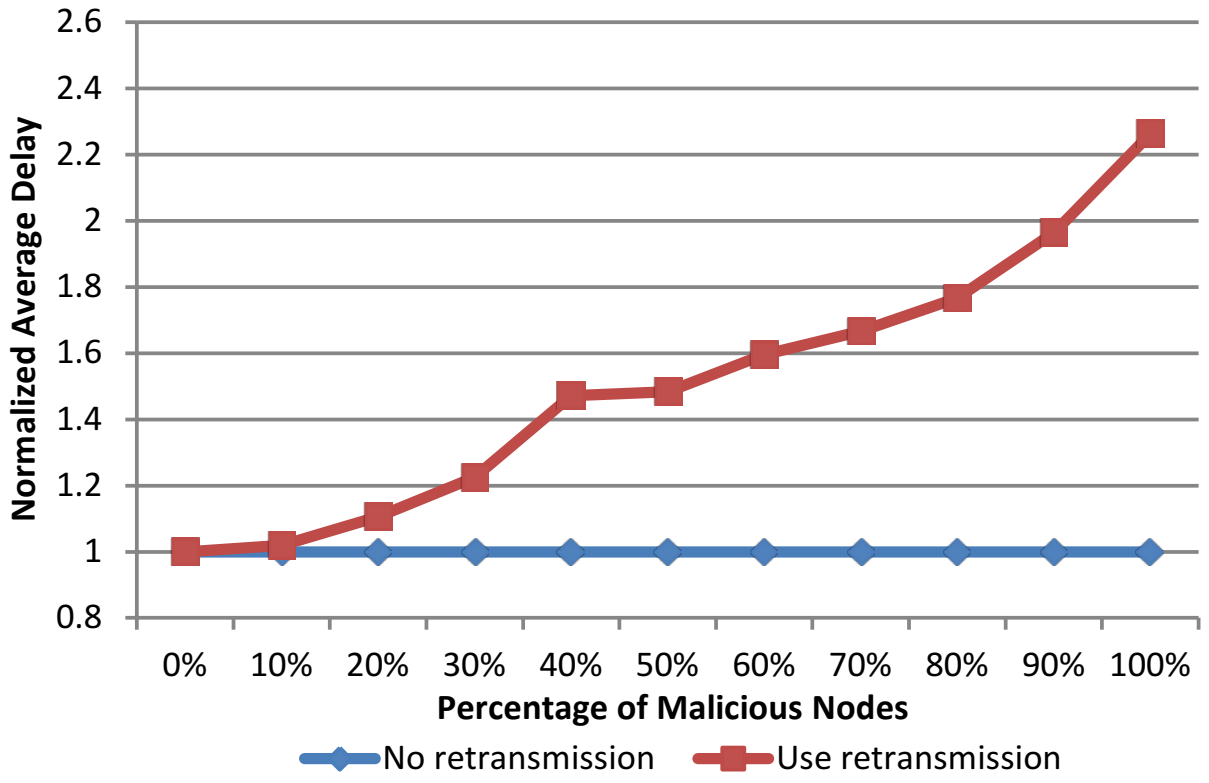

Fig. 6. Average delay increasing using retransmission.

In order to be consistent with the results of other approaches, we also looked at the impact on delivery ratio of using retransmission for 3000 randomly generated target profiles. Note that in this 3000 experiments, not all packets will get delivered in the scenario where no malicious node exists. $P_2$ and $P_3$ are also set to 80% and 60%, respectively.

In case of using retransmission, we compare the results between the scenarios where the malicious nodes are blocked or not. As shown in Figure 9, blocking out malicious nodes during gradient ascend phase show only very slight improve-ment in terms of delivery ratio. We believe it is due to the fact that the next node encountered (other than the malicious node who drops the packet) could also be a malicious node while the detected malicious node could possibly behave normally in its next encounter with the sender. The average delay, as shown in Figure 10, shows unpredictable pattern between these two scenarios, since blocking out malicious node does not have any impact on delay.

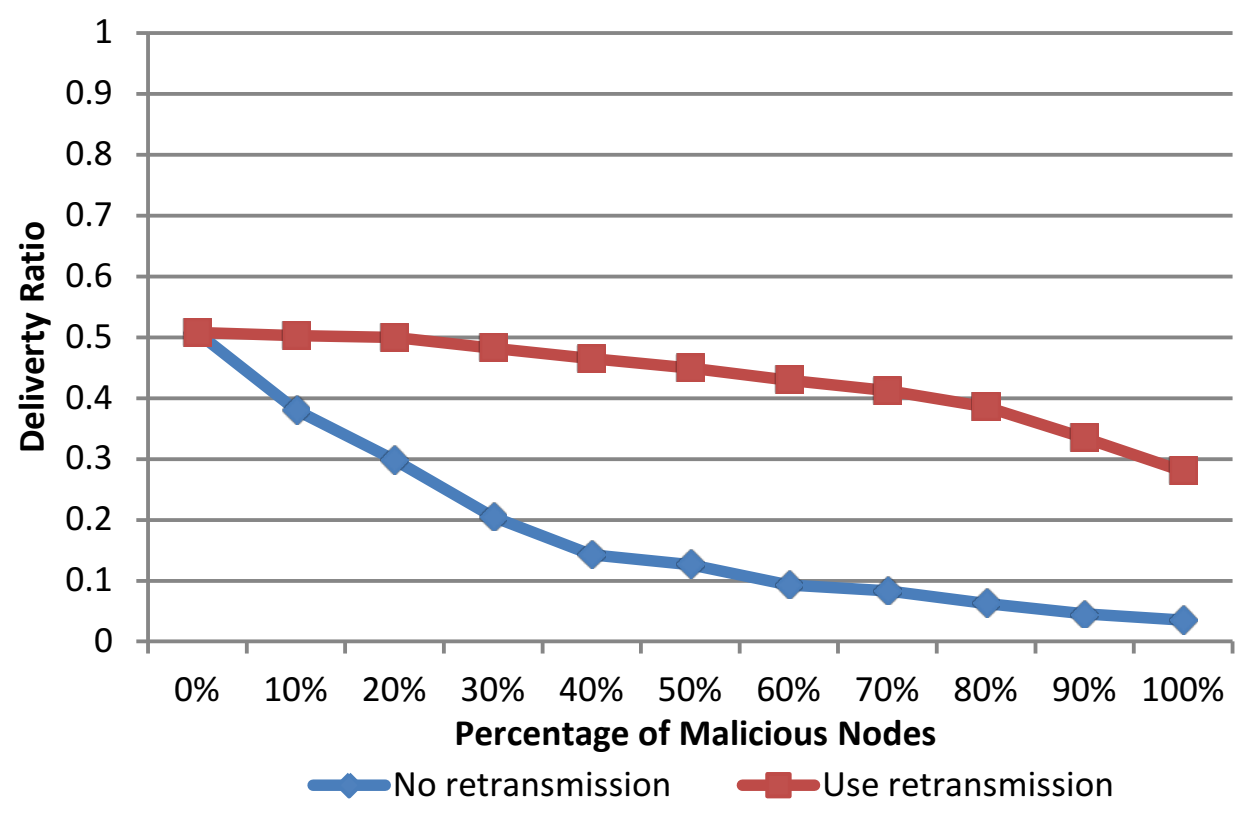

Fig. 7. Delivery ratio improvement using retransmission (3000 experiments).

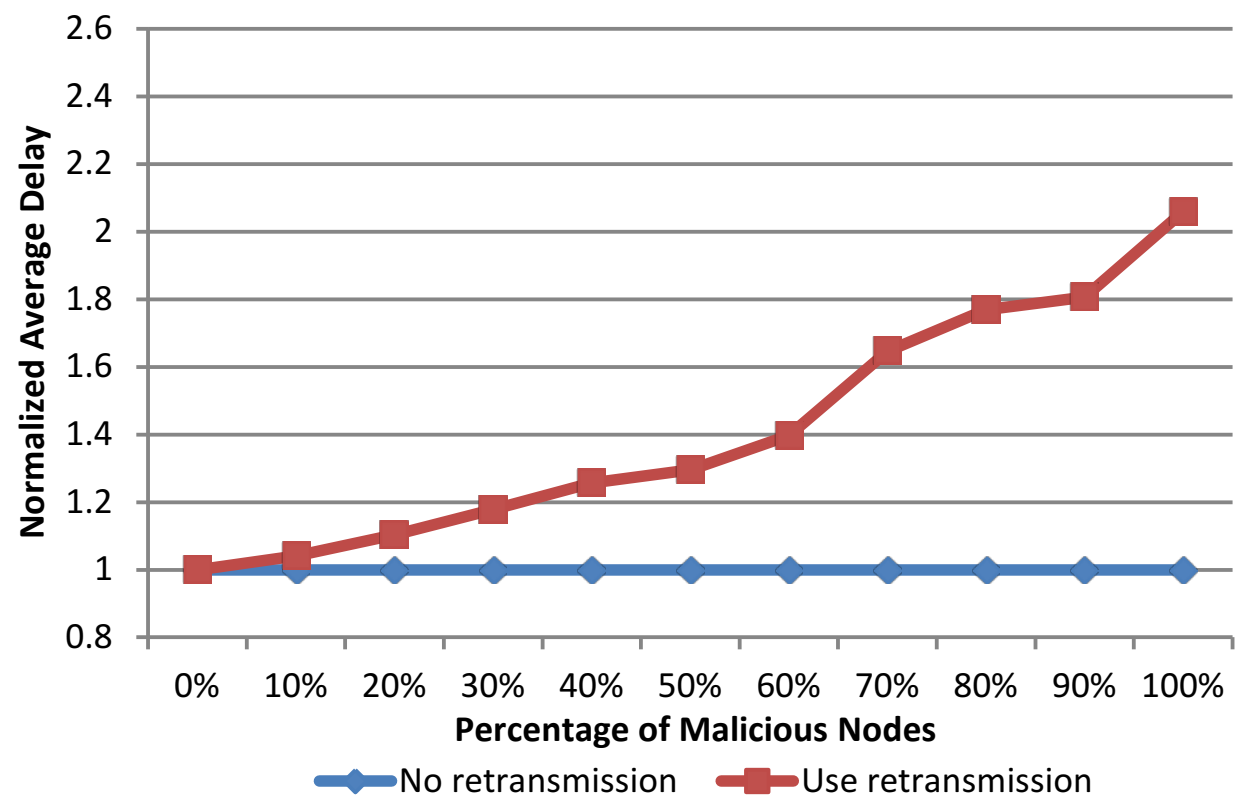

Fig. 8. Average delay increasing using retransmission (3000 experiments).

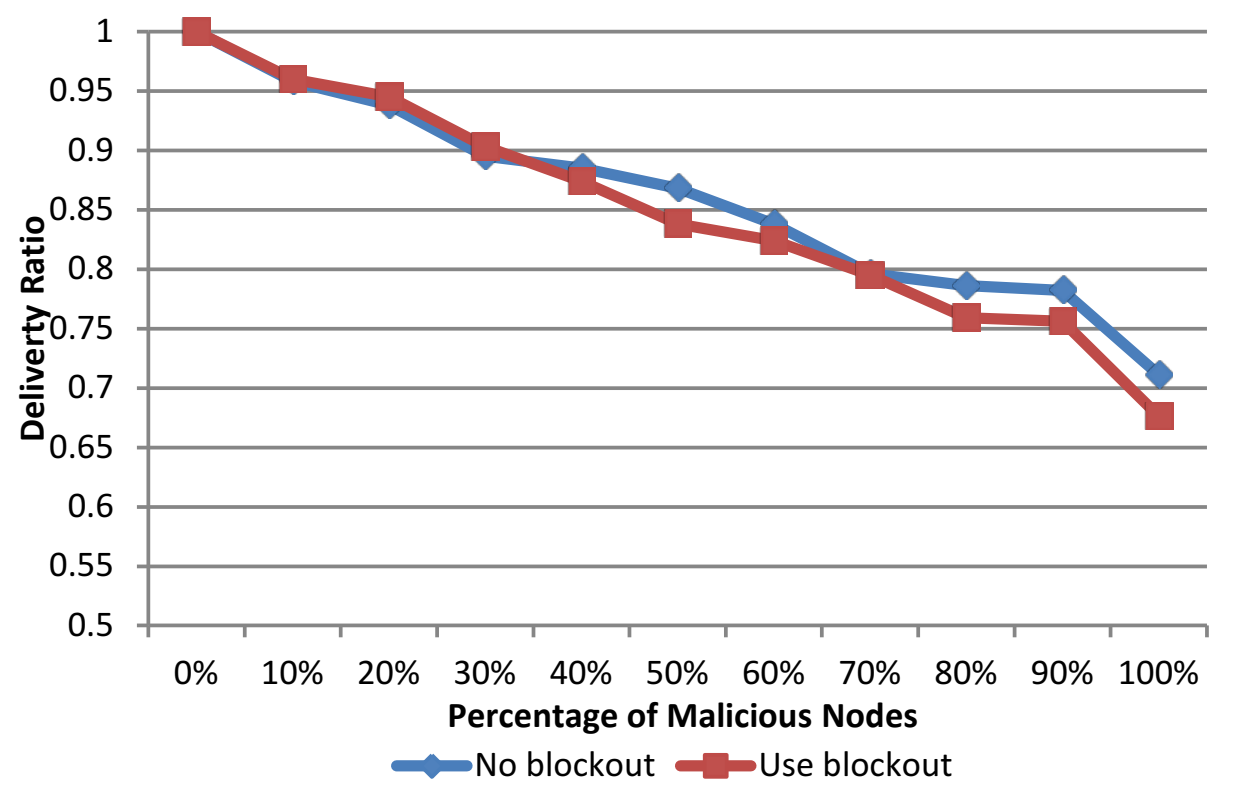

Fig. 9. Delivery ratio comparison with/without malicious blocking.

## VI. Credit based policing of malicious nodes

Credit based system for monitoring malicious nodes in an ad-hoc network was used by [3], [11]. Each node has some amount of credit, which is changed when it transmits or forwards a packet. The basic philosophy for the credit-based monitoring is as following.

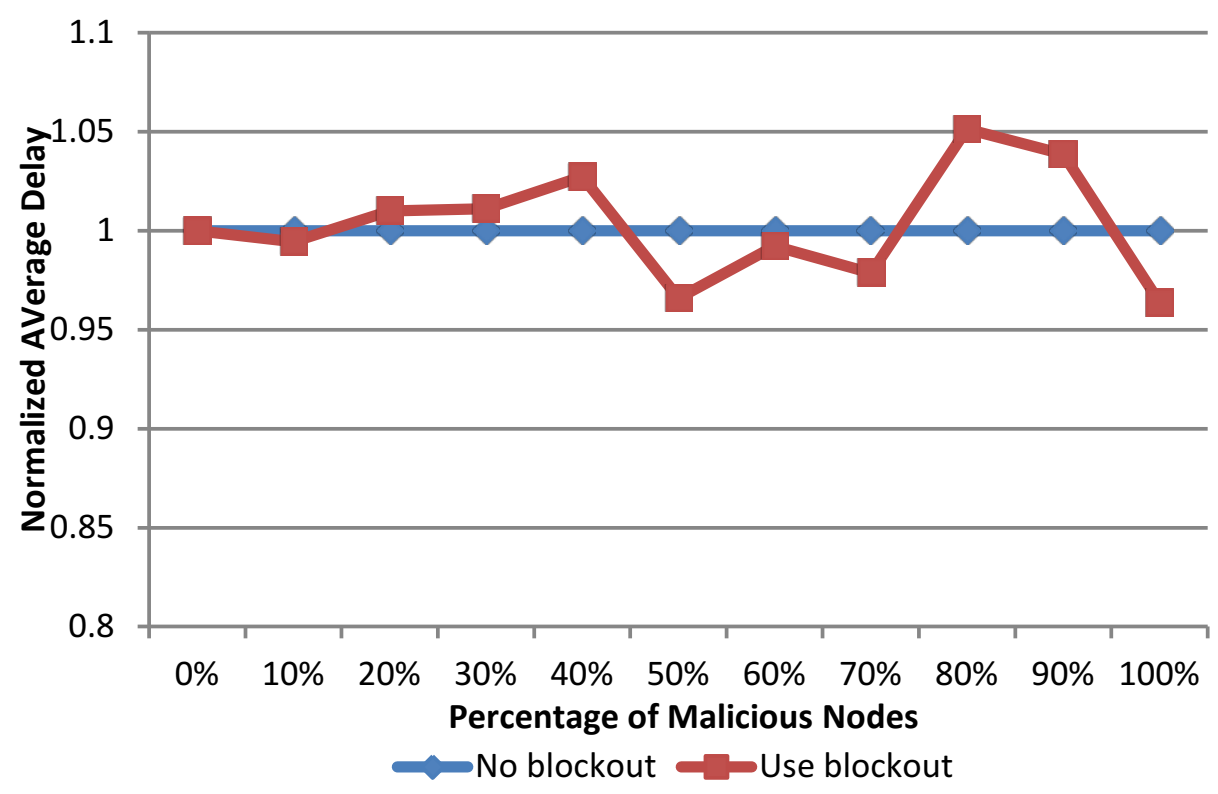

Fig. 10. Average delay comparison with/without malicious blocking.

1) When a node has its own packet to send, its credit gets decreased by the number of hops required to reach the destination. If the node does not have enough credit, it can not transmit the packet.
2) When a node forwards someone else's packet, its credit gets increased.
3) The credit on each node is monitored by either a centralized Credit Clearance Service (CCS) [3] or using a tamper-proof hardware [11], which allocate and delete credit from nodes as per rules.

This method effectively combats performance degradation in presence of malicious nodes in the network. For example, if a node acts maliciously for a long time and does not forward any packet, effectively no credit will be gained by it. As a result, when it wants to transmit its own packet, it will lack enough credit to successfully accomplish that. Hence, the node is forced to act normally and forward packets from others until it has enough credit to transmit its own packet. We have used the same scheme for Delay Tolerant Networks employing Profile-Cast. It should be noted that the chief concern here is the successful packet transmission ratio. In Section VI-B, we have compared this credit-based approach with the original scenario, where there are no monitoring of malicious nodes, to show how the packet transmission success rate is enhanced using this approach.

### A. Challenges

As we have seen in Section VI, a node loses its credit when it transmits a message. Let's call this loss amount as $credit_{th}$. Yale et al.[3] proposed a quantitative measure of $credit_{th}$ as the number of hops the packet from the sender takes to reach the destination. An example illustrating this is shown in Figure 11. Node 0 is the originating node, from where a packet gets transmitted. The packet has to travel $n$ hops before it reaches Node $n$, which is the destination. It can be seen that the Node 0, because of its interest for transmission has to lose a credit of $n$. On the other hand, the other nodes, which actively forward the packet to reach the destination, gains 1 credit each.

However, a major difference of using this value of $credit_{th}$ in Profile-Cast is that while Yale et al. [3] implemented their system in Mobile Ad-hoc Networks, Profile-Cast is in Delay-Tolerant Networks(DTNs). In DTNs, it's difficult to get a pre-conceived estimate of how many hops will be needed to route the packet finally to the destination. Hence, assigning the initial value of $credit_{th}$ based on the number of hops is not possible in this scenario. We have seen that in our experiments, in average about 4 hops are necessary to transmit a packet from the sender to the destination. Therefore, we have used for $credit_{th}$ a constant value of 4 that dictates the amount of credit loss when a node originates the transmission of a packet. Variation of the successful transmission ratio with $credit_{th}$ is shown in Section VI-B.

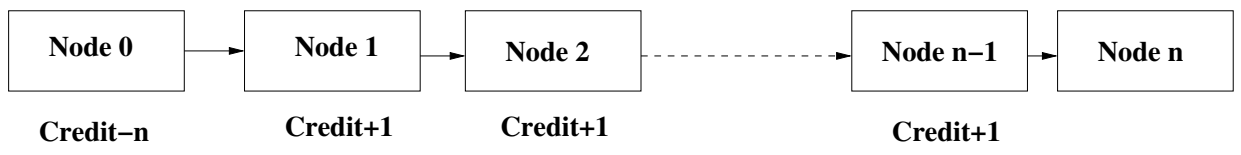

Fig. 11. Credit based monitoring scheme

Yale et al. [3] have assumed a lot of misbehaving scenario and proposed various solutions to them. These include collusion between the sender and the receiver, dropping of receipt and so on. For simplicity, we have neglected these scenario. Our only concern are the malicious nodes in the network that can drop packets based on the profile as described in Section IV. Due to scalability issues, we have not used a CCS (which was used by [3]) in our approach. Instead we have used a tamper-proof hardware that was used by [11]. The tamper-proof hardware, present in each node, stores its credit. The node can not, in any circumstances, modify the contents of the hardware, hence the name. Therefore, the credit check and updating is done in a distrbuted fashion inside each node, and not in the centralized server. Taking all these into consideration, we have devised our algorithm as shown in Algorithm 1.

**Algorithm 1** Credit based monitoring of malicious nodes

**if** Node $N$ wants to transmit a packet **then**
  **if** $Credit_N < credit_{th}$ **then**
    Transmit the packet;
    $Credit_N = Credit_N - credit_{th}$;
  **end if**
**end if**
**if** Node $N$ forwards a packet **then**
  $Credit_N = Credit_N + 1$;
**end if**

The malicious node adopts a strategy for behaving when this credit based scheme is applied. It has been seen from prior discussions, that malicious nodes can either act properly like any other normal nodes by forwarding packets, or can act maliciously by dropping packets. In the former case, it does not matter since the credit score gets increased. However, when acting maliciously, it is losing its credit which it could have gained by forwarding the packet. If the node continues to act maliciously, it will not gain enough credits and hence, after a certain time duration, the node will not have enough credit to start its own transmission. To prevent that, the malicious node, before misbehaving, check its credit value. If the credit of the node is less than $credit_{th}$, it does not misbehave, so that it can regain its credit lost in transmission. The procedure is shown in Algorithm 2

**Algorithm 2** Behaviour of malicious nodes

**if** Node $N$ is a malicious node **then**
  **if** $N$ does not act maliciously **then**
    Forward the packet;
  **else**
    **if** $Credit_N < credit_{th}$ **then**
      Forward the packet;
    **else**
      Drop the packet;
    **end if**
  **end if**
**end if**

In Section VI-B, we have fixed the two parameters $credit_{th}$ and the number of transmissions per experiment. Then, we have explored how the introduction of malicious nodes affect the success rate of transmission. Finally, we have compared the credit-based approach with the retransmission scheme used in Section V.

### B. Experiments

For effective use of credit-based scheme in profile-cast, we have modified the existing profile-cast simulator. In our experiments, we deal with the traces obtained from USC for the month of November 2007. Using the methods discussed in Section IV-C, we have reduced the traces to 3426 unique MAC addresses and 135 Access Points.

In order to implement the credit based approach in profile-cast, we have to first fix the value of $credit_{th}$[3]. As we have discussed earlier, it has been seen that the messages take an average of 4 hops to reach the destination. Therefore, the value of $credit_{th}$ should be around 4. We have conducted simulations with the $credit_{th}$ value varying from 3 to 5 in steps of 1, that is, we have oscillated around the 4 value. Each of these simulations have the same set of 3000 transmissions[4]. The number of malicious nodes is kept fixed at 20% for all the runs in this experiment. For all the experiments in this section, including this one, the malicious node misbehaving probability is kept between 80%-100%. The malicious node similarity threshold, as described in Section IV-C have been chosen as 60%. The flooding threshold is kept at 80%. The results have been shown in Figure 12. It can be seen that after the $credit_{th}$ value of 4, the performance improvement is very less. Therefore, we have chosen a $credit_{th}$ value of 4 for our next experiments.

In our next experiment, we vary the number of packets transmitted for each simulation to see the effect of credit based monitoring. We have varied the number of packets from 1000 to 3000 in steps of 1000. As before, we have assumed presence of 20% malicious nodes. The other parameters, that is, the malicious node misbehaving probability, the similarity and flooding threshold are kept at the same value as the last experiment. The $credit_{th}$ value is kept fixed at 4. The results have been shown in Figure 13. The *Original* column presents the results when no form of monitoring is present on the malicious nodes. We have used this convention for all the figures in this section. It can be seen that our credit based scheme gives an improvement in all cases, which is quite expected. The maximum benefit is obtained when the number of transmissions is 3000 and the improvement is found to be about 26%. For the next experiment, we have used this value, that is, 3000 transmissions so that the difference between the two approaches become visible enough.

[3]This denotes the minimum credit limit, beneath which, a node cannot start transmitting its packet

[4]These were randomly generated in the same way as described in Section IV-C

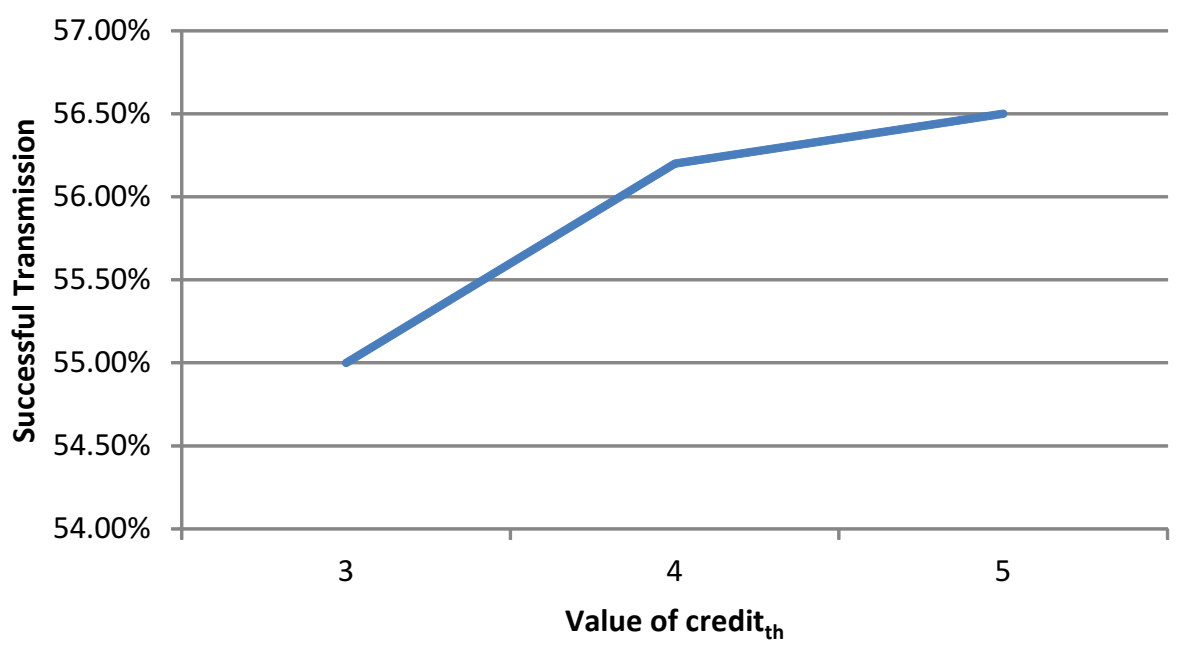

Fig. 12. Variation of Performance with $credit_{th}$

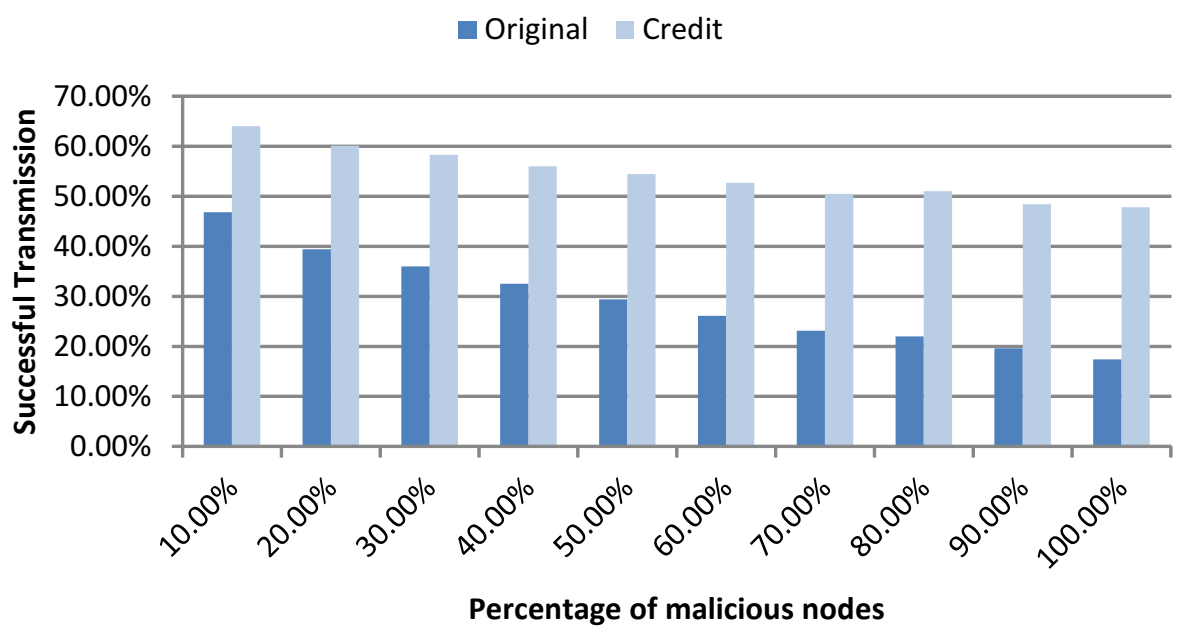

Fig. 14. Comparison of Performance with variation in number of malicious nodes

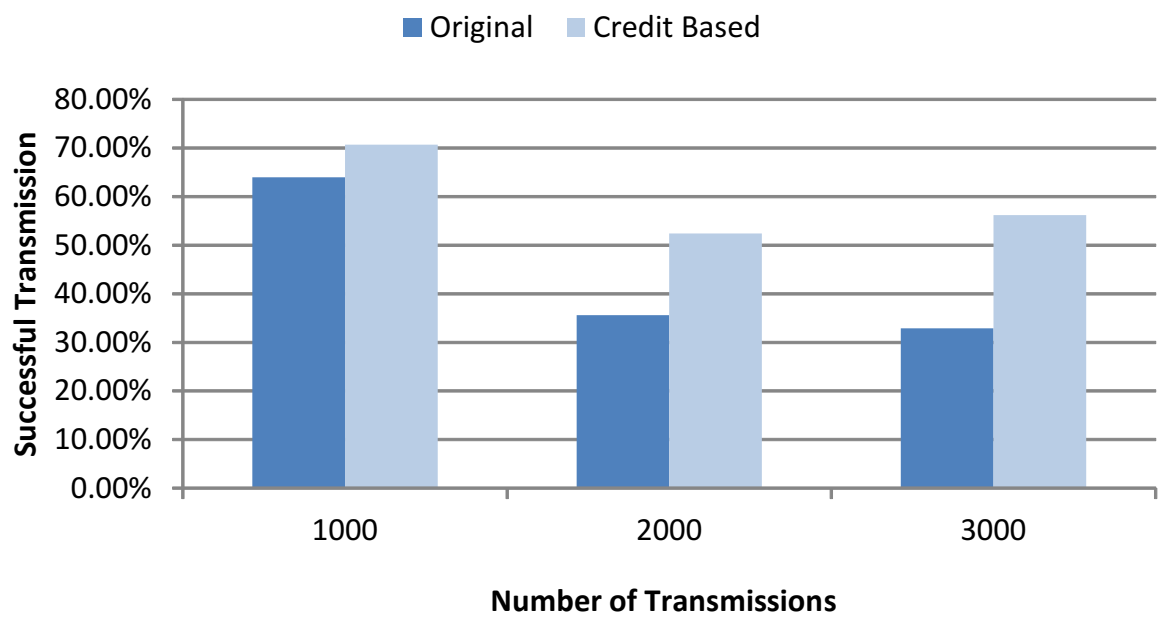

Fig. 13. Comparison of Performance with variation in number of transmissions

Now, we would like to explore the effect of credit based system when the number of malicious nodes are varied. Till now, all the experiments conducted had about 20% of malicious nodes. We now increase the number of malicious nodes from 10% to 100% in steps of 10%. The number of transmissions is kept at 3000 and the value of $credit_{th}$ is fixed at 4. As before, the other parameters are kept at the same value as the last experiment. The results have been shown in Figure 14. It can be seen that the improvement obtained through credit based scheme tends to increase when we enhance the number of malicious nodes, and it reaches about 30% when there are 100% malicious nodes. The reason is quite obvious. With more malicious nodes, the original method[5] shows a rapid decrease in performance, while for the credit based scheme, the decrease is much less due to the monitoring. Hence, the improvement, that is difference of successful transmission percentages between the two is more.

[5]One without any monitoring

Now, we compare our results with the retransmission scheme discussed in Section V. In this case, loss of packets are being countered using retransmission, which should effectively increase the successful transmission ratio. The results are shown in Figure 15. We have assumed presence of 20% malicious nodes and a $credit_{th}$ value of 4. The number of transmissions are varied from 1000 to 3000 in steps of 1000. As can be seen, the credit based scheme performs better than simple retransmission in case of 2000 and 3000 transmissions. However, for 1000 transmissions, it performs significantly worse than the retransmission scheme. The reason for this can be attributed to the fact that the credit based scheme takes some time to mature since unless some malicious nodes have actually lost credits by their own transmission, they will not act properly. Hence, for small transmission frames of 1000, retransmission schemes work better. The high delivery ratio for the experiment with 1000 nodes is because these 1000 experiments have been chosen intentionally such that, if there are no malicious nodes, the packets will definitely reach the destination. That is, there will not be any packet loss due to absence of similarity.

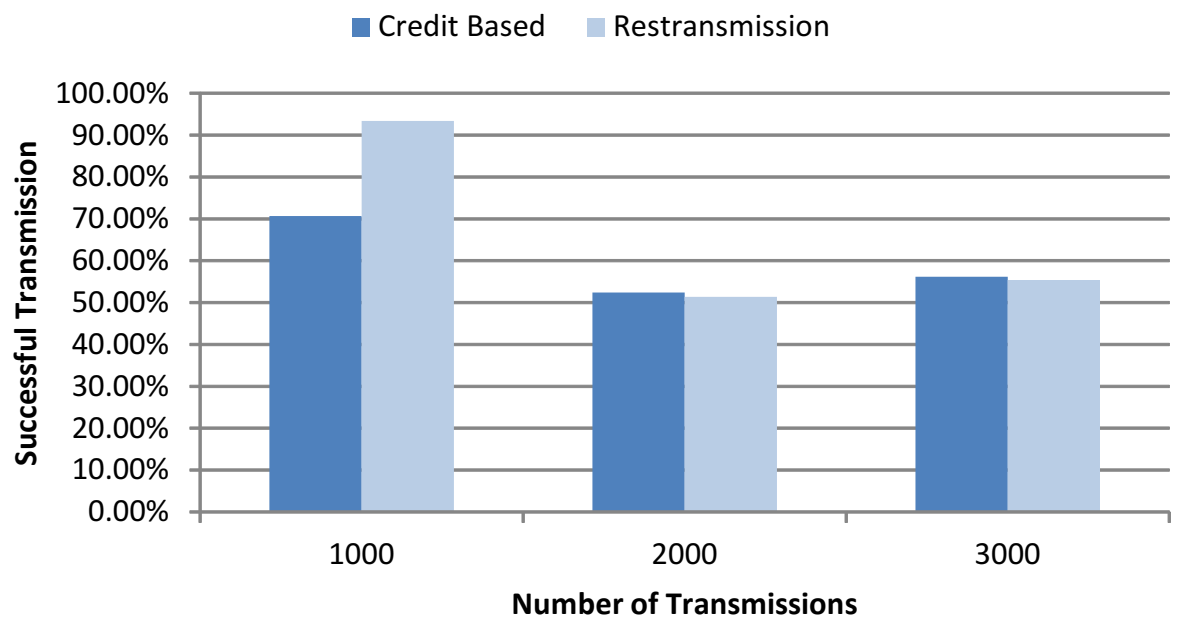

Fig. 15. Comparison of performances between credit based and restransmission scheme

Thus, the credit based monitoring of malicious nodes provides a significant improvement over the original scheme, where no monitoring was present. However, the additional burden that this scheme has to carry is the extra cost and overhead of the tamper-proof hardware, which monitors the credit in a distributed fashion.

## VII. An approach of self policing in behavioral aware network through reputation

CSI: Target mode of profile cast has two main phases: a) gradient ascend phase and b) group spread phase. In target mode, during the initial phase of gradient ascend, there will be very few nodes with very low profile matching. Since there are very few nodes who at least have some kind of matching, the sending node will have very less choice in-terms of selecting the next sender and since the matching of profile is very less, it is most likely that those nodes will not be very trust worthy as they may not be interested/acquainted about the target profile which the message is all about. On the other hand after the message has crossed the threshold similarity, there is a group spread phase and in this group spread phase the nodes carrying the message are very similar to the target profile. That is we can say that this message was targeted for them and based on this reasoning we can assume, the chances that these nodes within target profile will misbehave is extremely low and therefore we can ignore.

This new approach of reputation based self policing mechanism is based on this above mentioned reasoning. It tries to help the network learn about the presence of malicious nodes and helps it to evolve so that the impact of such malicious activities can be minimized during gradient ascend phase. This mechanism takes help of nodes in the target profile (on the boundary of the group spread phase to be specific) to gather information about successful delivery of the packet.

### A. Description of reputation based self policing mechanism

In this mechanism, the format of the message will be augmented to incorporate a footer section in the message. Each node will have an unique node $ID, NID$ (similar to some account number or email id) with exclusive access by the node. During gradient ascend phase of profile cast, after finding a node which is more similar than itself, the current message holder node will at first, append its own unique NID in the message footer and then deliver it to the new node. Thus during subsequent delivery of the message, all NIDs of the nodes who helped in carrying the packet will be appended in the message footer section. When, at the end of gradient ascend phase, the similarity of the nodes crosses a threshold towards the target profile, the node located at that flooding or group spread boundary will parse the message footer and send an exclusive message specific acknowledgement($ACK$) to each of the NIDs present in the footer.

Here, for the sending of acknowledgements we are relying on infrastructure network or the Internet. The assumption being each node who has an $NID$, at certain point of time in future will get associated with some access points (APs) and download the $ACK$ sent to its $NID$, by the node located on the flooding boundary, for a particular message, Notice that it has not been assumed that after each packet sent it will immediately expect an $ACK$.

This $ACK$ is very meaningful for the node, as it implies that what ever path was chosen to deliver the message did not have malicious nodes in it. On a local perspective, this $ACK$ means whichever neighbor it had chosen, based on similarity, to deliver the packet is not malicious (at least for that type of target profile).

Using this information, each node can maintain a table about its neighbors who are trustworthy for a certain type of profile. In this table the nodes will be classified based on type target profiles and there will be various "levels of trust. The neighbors who were not chosen to deliver a packet or for whom no acknowledgement was received will not enter in this table. There can be various levels of trusts, like if single acknowledgement is received for a node the level of trust is 1, if after that more acknowledgements are received where that node was again chosen as a sender, his trust value may increase to 2 and then 4 and then 8 etc, up-to a certain value after which it will be saturated and will be counted as most trusted neighbor. There should be an aging mechanism like: if for a very long time no $ACK$ is received for a node then his trust level will decrease gradually. This aging can be reasoned as follows: although it was one of the most trusted nodes but its current behavior is suspicious. This may happen if some good node was compromised at later point of time, and this aging process will help to avoid those nodes and incorporate a dynamic nature in this trust mechanism. Based on the above discussion the protocol of profile cast for choosing next node can be modified as shown in Algorithm 3.

**Algorithm 3** Modified Gradient Ascend

```
N = nearest neighbor in terms of time
delivery_factor = 0
while delivery_factor < c4 do
  if N.Similarity > ME.Similarity then
    Profile P = FIND_PROFILE
    if P.N.reputation = 8 then
      Add_to_footer (ME.NID)
      Deliver the message to N
      delivery_factor = delivery_factor + c1
    else
      if P.N.reputation >= 1 then
        Add_to_footer (ME.NID)
        Deliver the message to N
        delivery_factor = delivery_factor + c2
      end if
    else
      Add_to_footer (ME.NID)
      Deliver the message to N
      delivery_factor = delivery_factor + c3
    end if
  end if
  N = another neighbor
end while
```

Here $c1, c2, c3, c4$ are constants and is used to specify how many number of threads will be created or In other words how many copies of the message will be given to neighbors to avoid malicious activities. Certainly,

$$c4 = c1 > c2 > c3$$

These numbers can either be found through simulations or can be made as part of the learning process itself. Thus if the current node does not find any neighbor who has a good reputation, it will deliver more than one copies to its neighbors.

This reputation based scheme is certainly better than giving more than one copies of the packet each time, in terms of overhead because it tries to minimize the number of copies of the message by trying to find a reputed node in its neighborhood for that type of profile. Here node based on profile is chosen as follows, when a new message is received, its similarity is compared with previously received profiles and if this similarity is higher than a threshold then reputed nodes corresponding to that stored profile is considered. If there is no match with stored profiles, then after receiving the *ACK* corresponding to this message and corresponding to the neighbor selected to deliver this message, a new entry is created and this profile is saved and that node is also saved as trusted node for this profile. The number of such entries in the table can be constrained to avoid excessive storage overhead. It is easy to see that this mechanism helps the nodes to gradually learn about the network and about malicious activities. It tries to classify trustworthy nodes based on profiles and therefore handles both general malicious activities and profile based malicious activities. Therefore this mechanism has a "learning phase after which it is expected to perform better.

### B. Simulation

For simulation, existing code for profile cast simulator was modified. We have used USC November 2007 traces. The methods described in Section IV-C and Section VI-B are used to extract out 3426 unique MAC addresses and 135 *APs*. Then we randomly marked 50% nodes as malicious with 10% profile based misbehaving probability. Then we created 30 experiments where packets are transferred from source nodes to randomly generated target profiles. The number of hops to reach the target profile and the delay was noted. The plot of those two parameters are shown in Figure 16 and Figure 17 respectively.

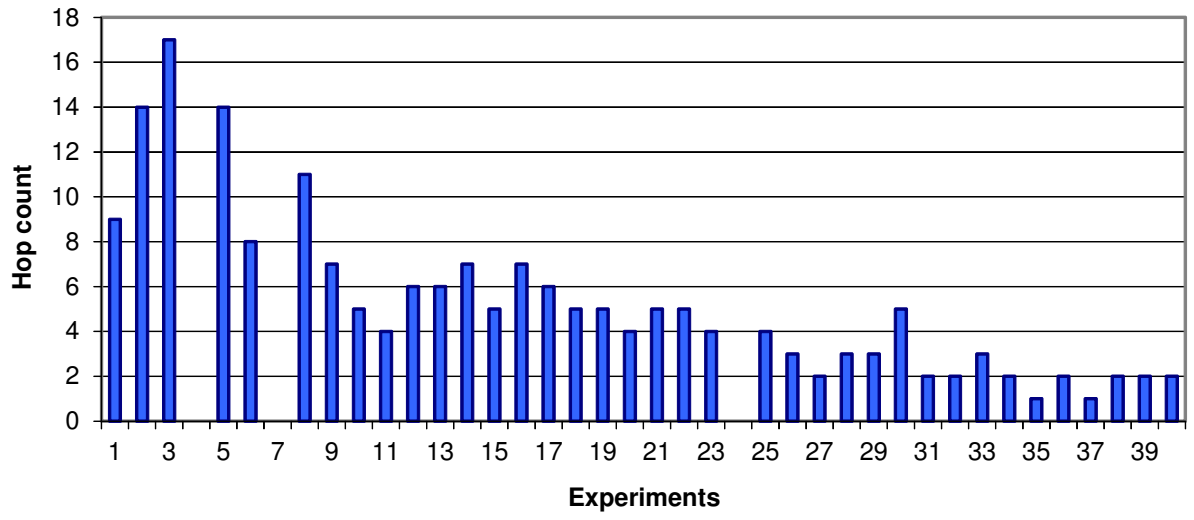

Fig. 16. Comparison of Number of hops needed to reach the target profile

It can be seen as with increasing index of experiments the network "learns" about presence of malicious nodes and tries to avoid paths through them and thus number of hops and delays get improved.

To measure how the delivery ratio decreases with percentage of misbehaving nodes, another 10 set of 1000 experiments were performed where percentage of misbehaving nodes were varied from 1

One major drawback of this scheme is it relies on infrastructured network to send *ACKs*, although that is a reasonable

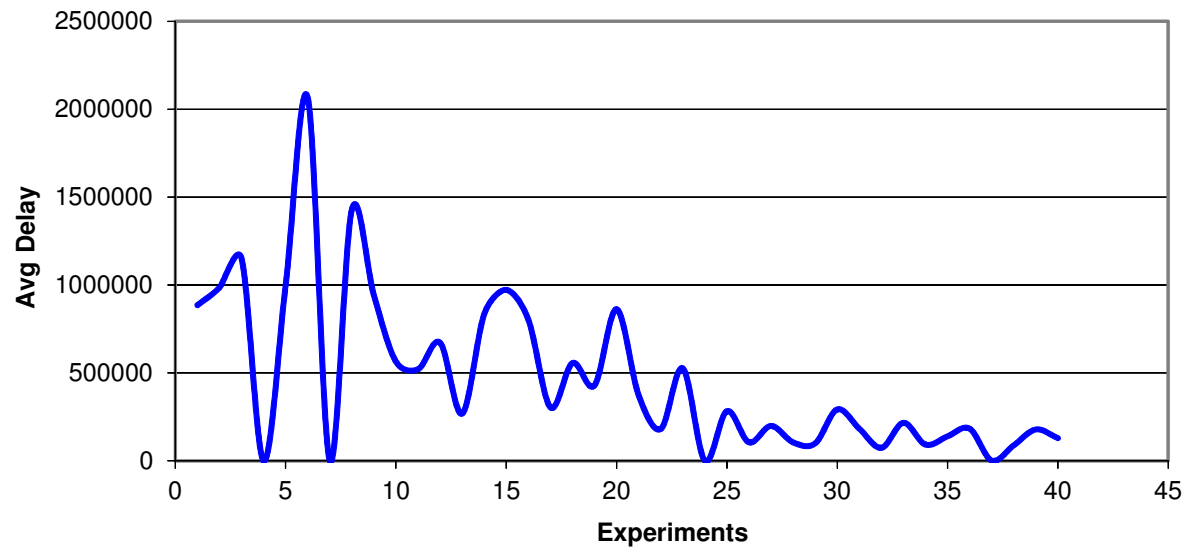

Fig. 17. Comparison of delay

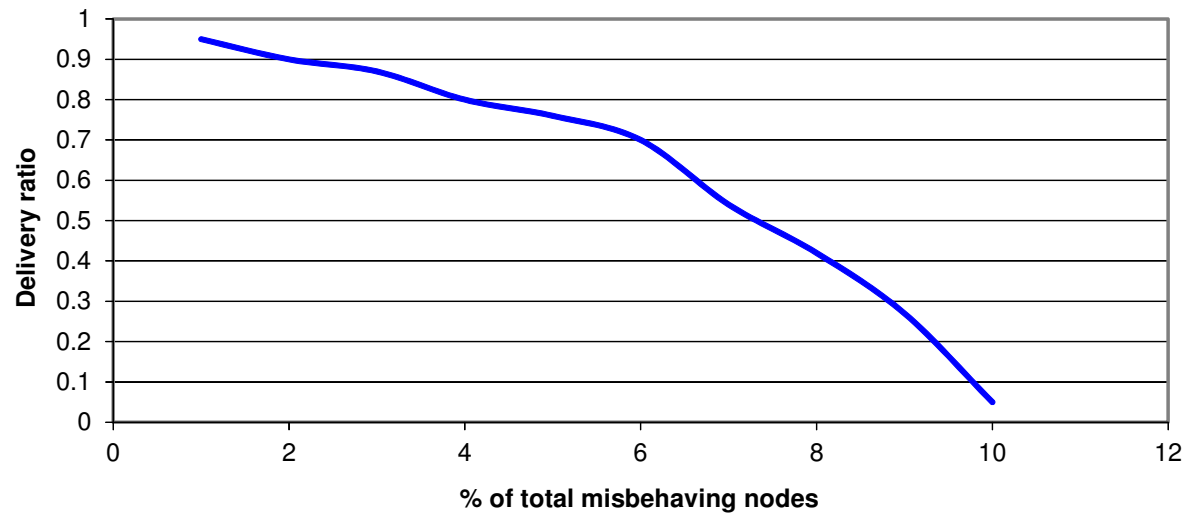

Fig. 18. Comparison of delivery ratio

assumption in campus like scenario where each node can be assumed to login at certain point of time and the profile information about nodes were itself gathered through access point traces. This draw back can be avoided in future work where sending the ACKs back through the same route can be judged. In general sending the *ACKs* back through the same nodes is not a practical idea in *DTN*, but since in this case the scenario is little bit different in the presence of theses profile informations, this can be feasible. During gradient ascend phase the packet got delivered to the target profile and this implies two things

1) There was overlap between each consecutive nodes in terms of profile and since these profiles are normally periodic in nature, there is a finite chance of meeting those nodes again at some point of time.
2) Since the packet was successfully delivered, this means at least in that case nodes did not behave maliciously for that message profile. So there is a chance that they will also help in delivering the *ACK* back.

Some other schemes like Game Theoretic approach can be employed here to make this reverse flow of *ACK* more robust.

## VIII. Game Theoritic Approach

Previously Iterated Prisoners Dilemma and Genetic Algorithm have been used in the papers by Komathy et al [12], [13] and by Seredynski [14]. Our approach also adopts the strategy which is most beneficial for optimizing its profit, which results in forwarding the packets most of the time. Hence slowly the malicious nodes evolve to become good nodes.

### A. Classical Prisoner's Dilemma

There are 2 criminals who are separately interrogated by the police to own up to their crimes. Depending on whether they chose to own up (defect) or keep silent (co-operate) independently they are given a penalty depending on his action and the action which is taken by the other criminal.

| Prisoner 2 / Prisoner 1 | Co-operate | Defect |
|---|---|---|
| Co-operate | 1 yr jail each | Go free-10 yrs jail |
| Defect | 10 yrs jail-Go free | 1 yr jail each |

TABLE I
OUTCOME ASSOCIATED WITH EACH ACTION OF THE PRISONER

| | |
|---|---|
| N2 | F |
| N3 | F |
| N4 | D |
| N5 | F |

TABLE II
FOR NODE 1: CONTAINS INFORMATION OF ALL OTHER NODES IN THE SYSTEM

### B. Iterated Prisoner's Dilemma

If 2 players play the Prisoners Dilemma more than once in succession then it is called Iterated Prisoners Dilemma. To reach Nash Equilibrium (where no player can benefit by changing his strategy given that the strategy adopted by the other players is constant) is has been seen that the optimal strategy is to play defect all the time.

### C. Application of Prisoner's dilemma to Profile-Cast

Each node will store 2 tables and a score. Table VIII-C has information about the Forwarding or dropping action of all the nodes in the system when it encountered them and wanted the encountered node to forward its packets. Table VIII-C contains information about all the other nodes in the system and what forwarding or dropping decision it took when the encountered node requested it for packet forwarding. Each node in addition to these tables also stores a cumulative score which is maintained over a duration of the total number of experiments.

### D. Assumptions

The assumptions we made are as follows.

1) The nodes are rational.
2) The system has a starting point where there is past information of the tables.
3) The score of each node is initialized to 0.
4) Only one packet forwarding request is sent from the Sender Node at a time.
5) We have not taken into consideration link characteristics of the channel.

### E. How the Tables and the Score is used

When a node receives a packet it either forwards the packet or misbehaves. Based on the past action of the immediate sender of the packet and the current forwarding dropping action of its own self it updates its score. The score updating takes place according to Table VIII-E.

| | |
|---|---|
| N2 | D |
| N3 | F |
| N4 | F |
| N5 | F |

TABLE III
FOR NODE 2: CONTAINS INFORMATION OF ALL OTHER NODES IN THE SYSTEM

| Receipient Node / Sender Node | Forward(past) | Drop(past) |
|---|---|---|
| Forward(present) | 4-4 | 3-0 |
| Drop(present) | 0-3 | 1-1 |

TABLE IV
SHOWING SCORE ASSOCIATED WITH EACH ACTION OF THE NODES

After the score updating next is the Table updating. Based on the forwarding behavior of the Current Node, its own Table VIII-C entry for Immediate Sender Node is updated. Similarly Immediate Sender Nodes Table VIII-C entry for the Current Node is updated based on its behavior. If the Current Node forwards the packet the probability of it misbehaving for the next time is decreased with the assumption that it will always try to maximize its profit and try to adopt the strategy which is the most beneficial to it, i.e. the forwarding strategy. Also assuming that the probability of malicious or selfish node misbehavior is being decreased with each forwarding action it can be shown that after a sufficiently large number of experiments the throughput of the network improves considerably.

### F. Strength and Weakness of the scheme

The strengths of the scheme are as follows.

1) It is scalable since only $2n$ information is being stored in the tables of each node.
2) The score is updated based on the basis of receiving or not receiving acknowledgements from the recipient, and is dependent on only one hop.
3) No node is isolated; Hence Bad nodes can also send and forward packets.
4) The malicious nodes evolve over time, since selfishness does not prove to be optimal, i.e. Nash Equilibrium can be reached if each node constantly forwards packets.

The weakness of the scheme is that it does not take into consideration the link characteristics. Hence if a node is dropped due to a noisy channel, the score and tables are updated erroneously. However this does not provide any wrong information about the probability of the node misbehaving.

### G. Results

We have conducted results on the USC traces for the month of November 2007, using 3000 transmissions. In the following two experiments, we want to notice the variation of delivery ratio with malicious nodes. For the first experiment in this section, including this one, the malicious node misbehaving probability is kept between 10%-100%. The malicious node similarity threshold, as described in Section IV-C have been chosen as 50%. The flooding threshold is kept at 60%. The results are shown in Figure 19.

In the next experiment, the malicious node misbehaving probability is kept between 80%-100%. The malicious node similarity threshold, as described in Section IV-C have been chosen as 60%. The flooding threshold is kept at 80%. The results are shown in Section 20.

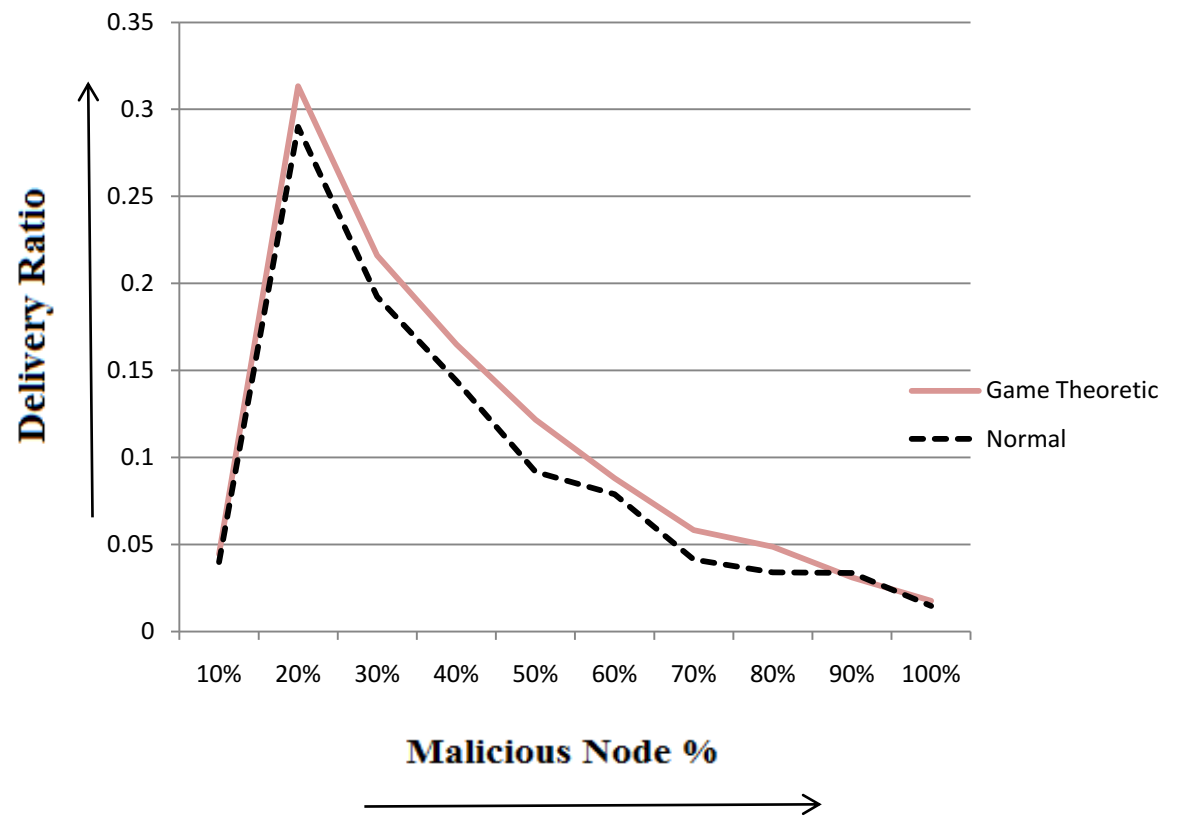

Fig. 19. Variation of Delivery ratio with the number of malicious nodes

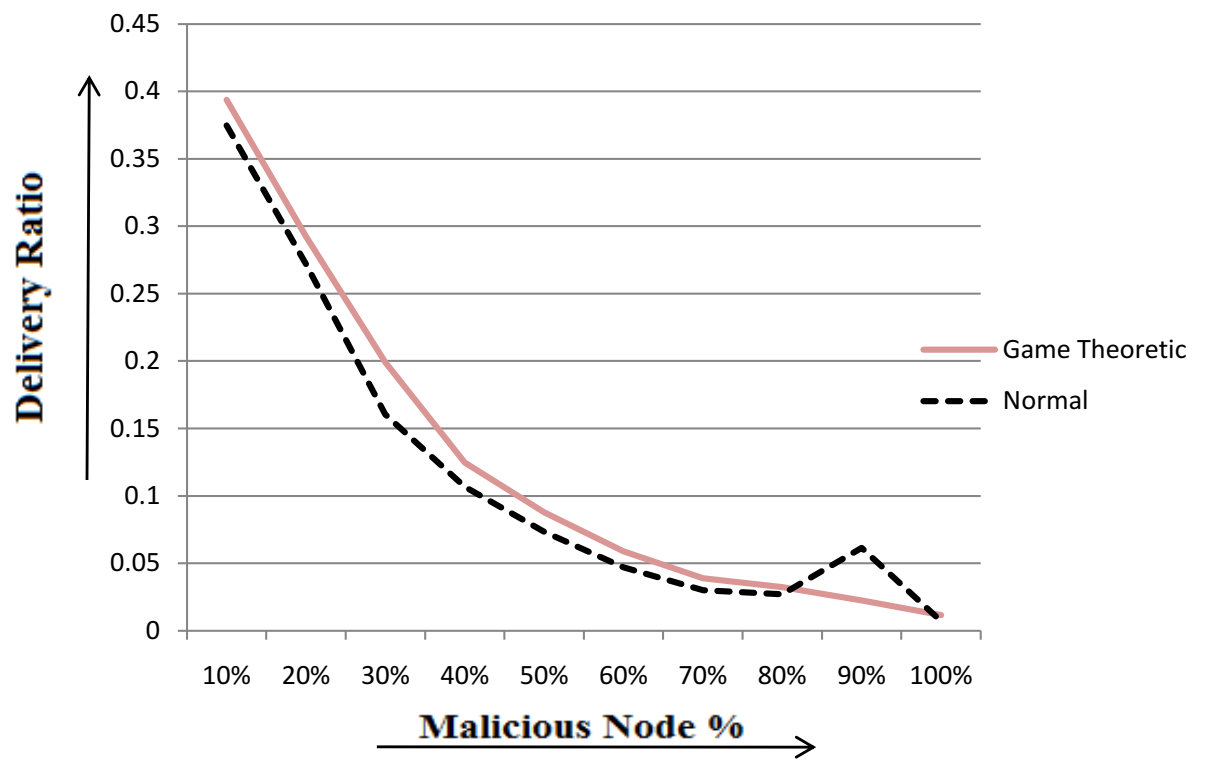

Fig. 20. Variation of Delivery ratio with the variation of malicious nodes

3 parameters have been varied for the 2 graphs. It shows that in majority of the cases the game-theoretic approach gives a better delivery ratio than the profile cast without self-policing. It can be inferred that if the number of experiments is more than 4000 the actual leap in performance can be experienced. Since here the number of unique nodes is 3426, while the number of experiments is just 3000. The improvement is not instantaneous as the network requires a long duration to reach a state of optimum performance where all nodes co-operate. There are a few deviation where the normal scenario shows better performance, This reason can be attributed to the fact that the path chosen by the packets select more good nodes than the game theoretic approach, or the selfish nodes do not misbehave in that particular experiment.

## IX. Future Work

For future work, we plan to investigate more alternatives for handling malicious nodes in behavior-aware mobile networks. For example, we can use multiple message copies in the network. Therefore, there will be multiple threads during the gradient phase instead of only one. In other words, duplication can be used to trade for performance and robustness. More subtle and well-tuned game-theory based approach is also worth the effort to investigate.